\documentclass[proceedings]{rmaa}

\title{Is the Plasma Within Bubbles and Superbubbles Hot or Cold? }

\author{Mordecai-Mark~Mac~Low
        \affil{Department of Astrophysics, American Museum of Natural
        History}}

\fulladdresses{
\item Mordecai-Mark Mac Low: Department of Astrophysics, American
        Museum of Natural History, 79th Street and Central Park West,
        New York, NY, 10024-5192, USA (mordecai@amnh.org)
}
\shortauthor{Mac Low}
\shorttitle{Bubble \& Superbubble Plasma}

\resumen{I review what is known about the temperature of the plasma
within stellar wind bubbles and superbubbles.  Classical theory
suggests that it should be hot, with characteristic temperatures of
order a million degrees.  This temperature should be set by the
balance between heating by the internal termination shocks of the
central stellar winds and supernovae, which expand at thousands of
km~s$^{-1}$, and cooling by conductive evaporation of cold gas off the
shell walls.  However, if the hot interior gas becomes dense enough
due to evaporation or ablation off of interior clouds, it will cool in
less than a dynamical time, leading to a cold interior.  The
observational evidence appears mixed.  On the one hand, X-ray
emission has been observed from both stellar wind bubbles and
superbubbles.  ON the other hand, no stellar wind bubble or
superbubble has yet been observed emitting at the rate predicted by
the classical theory: they are either too faint or too bright, by up
to an order of magnitude.  Alternate explanations have been proposed
for the observed emission, including off-center supernova remnants
hitting the shell walls of superbubbles, and residual emission from
highly-ionized gas out of coronal equilibrium.  Furthermore, the
structures of post-main sequence stellar wind bubbles, expanding into
what are presumably old stellar wind bubbles, appear in at least some
cases to show that the bubble interior is cold, not hot.  (The
classical example of this is NGC~6888.)  What is the actual state of
bubble and superbubble interiors?}

\abstract{I review what is known about the temperature of the plasma
within stellar wind bubbles and superbubbles.  Classical theory
suggests that it should be hot, with characteristic temperatures of
order a million degrees.  This temperature should be set by the
balance between heating by the internal termination shocks of the
central stellar winds and supernovae, which expand at thousands of
km~s$^{-1}$, and cooling by conductive evaporation of cold gas off the
shell walls.  However, if the hot interior gas becomes dense enough
due to evaporation or ablation off of interior clouds, it will cool in
less than a dynamical time, leading to a cold interior.  The
observational evidence appears mixed.  On the one hand, X-ray
emission has been observed from both stellar wind bubbles and
superbubbles.  ON the other hand, no stellar wind bubble or
superbubble has yet been observed emitting at the rate predicted by
the classical theory: they are either too faint or too bright, by up
to an order of magnitude.  Alternate explanations have been proposed
for the observed emission, including off-center supernova remnants
hitting the shell walls of superbubbles, and residual emission from
highly-ionized gas out of coronal equilibrium.  Furthermore, the
structures of post-main sequence stellar wind bubbles, expanding into
what are presumably old stellar wind bubbles, appear in at least some
cases to show that the bubble interior is cold, not hot.  (The
classical example of this is NGC~6888.)  What is the actual state of
bubble and superbubble interiors?}

\keywords{stellar wind bubbles --- superbubbles --- X-ray emission ---
OB stars}

\begin{document}

\maketitle

\section{Bubble Structure}

\begin{figure}[bth]
\begin{center}
\includegraphics{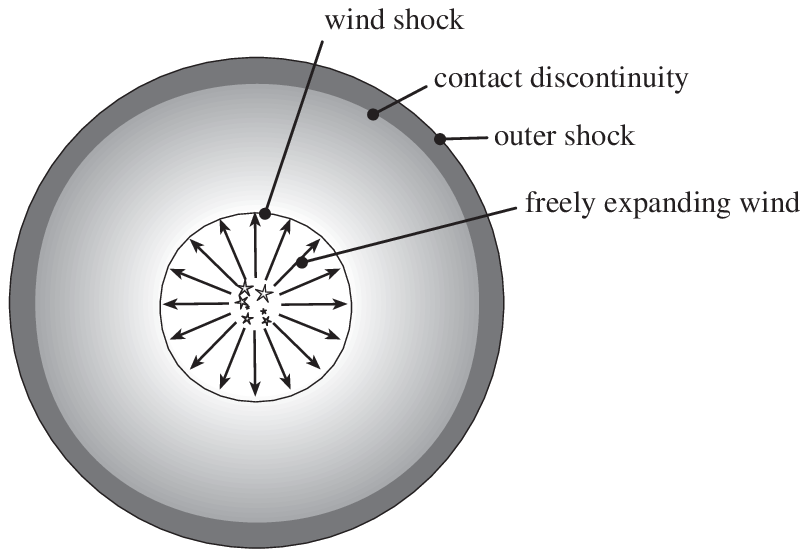}
\end{center}
\vspace{0.3in}
\caption{Structure of a classical bubble driven by continuous input of
mechanical energy from a central star or star cluster.  Diagram shows
inner and outer shocks, separated by a contact discontinuity across
which mass flows driven by conductive evaporation. \label{structure}}
\end{figure}

The interpretation of observations of plasma within bubbles and
superbubble relies on understanding the density and temperature
structure over which the observations integrate.  The touchstone for
such understanding remains the evaporative wind-blown bubble first
described by \nocite{cmw75}Castor, McCray, \& Weaver (1975) and
\nocite{w77}Weaver, et al.\ (1977).  Figure~\ref{structure} shows the
two-shock structure that results from continuous mechanical energy
input into a uniform medium.  The freely expanding wind is shocked at
the inner shock, heating it up.  The resulting pressurized region
drives a shock into the external gas, sweeping up a shell that is
usually dense enough to cool.  The functional form of the radius of
the dense shell $R$ can be derived from dimensional arguments by
noting that the only other physical variables in the problem are the
time $t$, the mechanical luminosity $L$, and the external number density
$n_0$.  Only one dimensionless constant can be assembled from
these, showing that 
\begin{equation} 
R \propto L^{1/5} n_0^{-1/5} t^{3/5}
\end{equation}
Thermal conduction occurs across the contact
discontinuity separating the hot interior from the cold shell,
evaporating mass into the interior.  The temperature $T$ and number density
$n$ in the interior can be derived from similarity solutions to
have functional forms $T(r) \propto (1-r/R)^{2/5}$ and $n(r)
\propto (1-r/R)^{-2/5}$ (Weaver et al.\ 1977)

At late times, radiative cooling can become important to the interior
of a bubble.  If the density in the interior follows the similarity
solution given above, then the cooling time can be approximated by
taking the cooling rate $\Lambda = (10^{-22} \mbox{ erg s}^{-1} \mbox
{ cm}^3) \zeta T_6^{-07}$, where $T = (10^6 \mbox{ K})T_6$ and $\zeta$
is the metallicity compared to solar.  The cooling rate for the bubble
is then given by $L = \int n^2(r) \Lambda(T(r)) d^3r$, and the cooling
time is
\begin{equation} \label{cool}
t_c = (16 \mbox{ Myr}) L_{38}^{3/11} n_0^{-8/11} \zeta^{-35/22}
\end{equation}
\nocite{mm88}(Mac Low \& McCray 1988), where $L = (10^{38} \mbox
{ erg s}^{-1}) L_{38}$.  Weaver et al.\ (1977) computed cooling rates
including a non-equilibrium treatment of the ionization structure of
the interior, and found cooling rates compatible with this result, as
shown in their Figure~6.  Note that, for typical external densities,
the cooling times are longer than the lifetimes of massive stars, so
for individual stellar wind bubbles cooling will not be very important
in the classical picture described here.

The question of thermal conduction across the interface has been
considered extensively.  The physical mechanism acting is that fast
electrons from the hot interior can penetrate significant distances
into the cold shell before depositing their energy in collisions with
the gas, transferring heat across the contact discontinuity.  The
resulting heating raises the pressure of the inner edge of the shell,
which then expands into the hot interior.  This heat conduction
saturates due to the electric fields set up by the movement of the
electrons \nocite{cm77}(Cowie \& McKee 1977).

Tangled magnetic fields are often invoked to suppress conduction.  The
idea here is that electrons tied to tangled field lines will have long
pathlengths without travelling very far into the cold shell, reducing
the efficiency of thermal conduction.  However, magnetic fields cannot
actually be tangled very much, as magnetic pressure and tension will
climb as they become more tangled.  \nocite{t95}Tao (1995) and
\nocite{ps96}Pistinner \& Shaviv (1996) have shown that tangling can
suppress thermal conduction by at most an order of magnitude. (Note
also that \nocite{b97}Boroson et al.\ [1997] have observed evidence of
conductive evaporation as discussed below at the end of \S~\ref{swb}.)
This will make a difference; however \nocite{sc93}Slavin \& Cox (1993)
studied the effect of reduced conduction in SNRs and found that,
although the interiors are indeed hotter, even small amounts can lead
to effective cooling in the end.

\section{Observations} 

If the cooling time given by equation~\ref{cool} is within a factor of
three of being correct then bubbles and superbubbles should be filled
with hot gas that emits X-rays.  The theoretical spectrum of this
X-ray emission, including the effects of non-equilibrium ionization,
was already modeled by Weaver et al.\ (1977).  After the launch of
{\em Einstein}, \nocite{bl85}Bochkarev \& Lozinskaya (1985) used the
Weaver et al.\ temperature and density profiles along with an
equilibrium ionization model to compute the expected X-ray luminosity
of a stellar wind bubble, applying the model to the Wolf-Rayet bubble
NGC~6888. .

NGC~6888 was indeed detected by \nocite{k87}K\"ahler, Ule, \& Wendker
(1987) using {\em EXOSAT} and by \nocite{b88}Bochkarev (1988) using
{\em Einstein}, but with much an order of magnitude lower X-ray
luminosity than originally predicted.  Inspired by this,
\nocite{bz90}Bochkarev \& Zhekov (1990) included the effects of
non-equilibrium ionization into a computation of the X-ray luminosity,
showing that it would be expected to reduce the luminosity.  An
alternative explanation of the unexpectedly low luminosity was offered
by \nocite{gm95a}Garc\'{\i}a-Segura \& Mac Low (1995) who took into
account the post-main sequence evolution of the central Wolf-Rayet
star, and used the shell dynamics to predict the X-ray luminosity, as
further explained in the following section.

The X-ray luminosity expected from superbubbles was computed assuming
equilibrium ionization and the Weaver et al.\ (1977) interior profile
by \nocite{cm90}Chu \& Mac Low (1990; typos corrected in
\nocite{c95}Chu et al.\ 1995).  They used a constant X-ray emissivity
of $\Lambda_x(T) = (3 \times 10^{-23} \mbox{ erg cm}^3 \mbox{ s}^{-1})
\zeta$ for $T > 5 \times 10^5$ K \nocite{rs77}(Raymond \& Smith 1977),
to find
\begin{equation} \label{lx}
L_x \simeq (8 \times 10^{27} \mbox{erg s}^{-1}) \zeta I(\tau)
n_0^{10/7} R_{\rm pc}^{17/7} v_{\rm km}^{16/7}, 
\end{equation}
where $R_{\rm pc}$ is the shell radius measured in pc, and $v_{\rm km}$ is
the shell expansion velocity measured in km s$^{-1}$, both of which
can be observed.  The ambient number density can be estimated using
the emission measure through the shell, with some assumptions, as
explained in the cited papers.

The effect of a non-uniform interior temperature and density on
spectral fits was investigated by \nocite{ss98}Strickland \& Stevens
(1998) using non-equilibrium ionization and the actual {\em ROSAT}
response function but neglecting thermal conduction in a numerical
computation (some conduction occurred anyway due to numerical
dissipation).  Although their computation probably did not have an
accurate internal structure because of the neglect of thermal
conduction, their main point was to emphasize how badly astray one can
be led by simple one or two temperature fits to an X-ray spectrum
coming from a complex structure.  Metallicities more than an order of
magnitude too low can be derived with such fits, for example.

\section{Stellar Wind Bubbles}  \label{swb}

As it turns out, the only stellar wind bubbles that have actually been
observed in the X-ray are two Wolf-Rayet ring nebulae, NGC~6888
(K\"ahler et al.\ 1987; Bochkarev 1988; \nocite{www94}Wrigge, Wendker,
\& Wisotzki 1994) and S~308 \nocite{w99}(Wrigge 1999).  Main sequence
bubbles tend to be too large and dim to be potentially observable by
the instruments available up until the latest generation of
large-aperture, X-ray telescopes, just as their shells are often too
dim to be observed in the optical \nocite{mvl84}(McKee, Van Buren, \&
Lazareff 1984).  Nevertheless, the observed post-main sequence bubbles
may offer revealing insights into the state of the interiors of main
sequence bubbles, as well as posing interesting problems themselves.

Stellar wind bubbles around Wolf-Rayet stars must be
interpreted taking into account the mass-loss history of the star
during its post-main sequence evolution. \nocite{d92}D'Ercole (1992) and
\nocite{gm95a}Garc\'{\i}a-Segura \& Mac Low (1995a) summarize the
basic idea, which was explored in more detail by
\nocite{gm95b}Garc\'{\i}a-Segura \& Mac Low (1995b) and
\nocite{gml96}Garc\'{\i}a-Segura, Mac Low, \& Langer (1996).  While a
massive star remains on the main sequence it blows a main sequence bubble
having roughly the structure described in the introduction.  It then
evolves into a red supergiant with a slow, massive wind that expands
into the cavity left by the main sequence bubble, possibly still
filled with high-pressure hot gas from the shocked main sequence
wind.  After the star loses its atmosphere, it will evolve back to a
blue Wolf-Rayet star with a high-velocity, low-density wind that
sweeps up the red supergiant wind.

This new stellar wind bubble evolves not in a uniform background
medium, but in the $r^{-2}$ density distribution of the red supergiant
wind.  This enhances the strength of the \nocite{v83}Vishniac (1983)
instabilities in the swept-up shell of red wind.  At the edge of the
dense red supergiant wind, the shell accelerates as it suddenly
blows out into the low-density main-sequence bubble.  The density
enhancements already produced by the Vishniac instabilities are then
strongly amplified by Rayleigh-Taylor instabilities, fragmenting the
shell.  NGC~6888 appears to be at just this stage of development,
while S~308 appears to be just at the beginning of the blowout
\nocite{c82}(Chu et al.\ 1982).

The morphology of the blowout can be used to probe the temperature of
the gas in the fossil main-sequence bubble as well.  If that gas is
hot, the expansion velocity of the shocked Wolf-Rayet wind blowing out
past the fragmented shell is subsonic, so it merely drives a sound
wave into the surrounding hot gas, generating no observable density
enhancements.  On the other hand, if the surrounding gas is cold, the
expansion of the hot Wolf-Rayet wind is supersonic, so it drives a
strong shock into the gas, compressing it and generating an observable
signature.  Garc\'{\i}a-Segura \& Mac Low (1995b) demonstrate this
effect in their numerical models (though the figure must be examined
carefully to understand this due to the dynamic range covered).  Over
the short lifetime of the massive progenitor of the central Wolf-Rayet
star of NGC~6888, the hot shocked wind in its bubble should not have
had the chance to undergo significant radiative cooling.  However,
observations of NGC~6888 in [OIII] \nocite{m89,m90}(e.g. Mitra
1989, 1990) clearly show strong filaments bounding the nebula a few
tens of arcsec outside of the fragmented shell, indicative of a strong
shock expanding into the surrounding medium, and thus of a cooled
main-sequence bubble, contrary to expectation.

The X-ray luminosity, combined with the shell dynamics revealed by
imaging and spectroscopy in emission lines such as H$\alpha$, yield
additional puzzles.  Garc\'{\i}a-Segura \& Mac Low (1995a) show that
the radius, velocity, and shell properties of NGC~6888 can be
consistently explained only if the mechanical luminosity of the
central star is roughly an order of magnitude lower than the generally
accepted value.  Clumping in the wind could give a factor of three
lower value \nocite{mr94}(Moffat \& Robert 1994), but the last factor
of three remains mysterious.  Using the lower mechanical luminosity
and the observed shell parameters, Garc\'{\i}a-Segura \& Mac Low
(1995a) are able to recover the observed X-ray luminosity of NGC~6888
(K\"ahler et al. 1987, Bochkarev 1988, Wrigge et al.\ 1994) from their
model as well.  The other Wolf-Rayet wind nebula detected in the
X-ray, S~308, shows the same inconsistency between shell dynamics and
accepted wind parameters.  In this case, however, the observed X-ray
luminosity is even fainter than the value predicted
\nocite{w99}(Wrigge 1999).

Causes for this enhanced cooling could include stronger conductive
evaporation than generally assumed, and higher metallicities in the
cooling gas.  Enhanced conductive evaporation may well occur due to
the increased surface area and mixing produced by \nocite{v83}Vishniac
(1983) instabilities (\nocite{mn93}Mac Low \& Norman 1993,
\nocite{gml96} Garc\'{\i}a-Segura et al.\ 1996).  In fact, evidence
for the existence of conductive evaporation has been observed in S~308
by \nocite{b97}Boroson et al.\ (1997).  They used the GHRS on HST to
observe the C~{\sc iv} and N~{\sc v} resonance absorption lines in the
spectrum of the central star of S~308.  The N~{\sc v} line, in
particular, was extremely broad (over 50 km~s$^{-1}$ FWHM), as
predicted by models of conductive evaporation.  Its strength was
rather greater than predicted by those models, however, suggesting
that the interior of this Wolf-Rayet bubble was enhanced in nitrogen.
This is not unexpected, as the interior mass should consist primarily
of mass evaporated off the cold shell of mass swept-up from a previous
red supergiant wind, which should indeed be enriched in fusion
products such as nitrogen.

\section{Superbubbles}

X-ray emission has indeed been observed from young superbubbles both
in our own Galaxy and in the Large Magellanic Cloud (LMC), as well as
from the even larger bubble structures blown from starburst galaxies
\nocite{s94,s96}(e.g.\ Suchkov et al.\ 1994, 1996).  The LMC
superbubbles are easier to analyze as they lie at known distances in a
roughly face-on disk with little obscuration along the line of sight.
The resolution of {\em ROSAT} and even of {\em Einstein} was high
enough to resolve even smaller superbubbles, allowing the correlation
of emission from cold H~{\sc i} and warm ionized gas emitting in
H$\alpha$ with hot, X-ray emitting gas.

The X-ray emission from LMC superbubbles can be compared directly to
the values predicted from equation~\ref{lx} for a Weaver et al.\
(1977) bubble using measurements of the shell dynamics taken from
long-slit or Fabry-Perot spectroscopy in optical emission lines such
as H$\alpha$.  Using such techniques, Chu \& Mac Low (1990) and Wang
\& Helfand (1991) showed that some LMC superbubbles were brighter
than predicted.  That this is not a universal property of the LMC
superbubbles was shown by \nocite{c95}Chu et al.\ (1995), who
described {\em ROSAT} observations of other LMC superbubbles with
roughly equivalent stellar content and dynamics that have upper limits
on their X-ray luminosity consistent with the Weaver et al.\ (1977)
prediction. 

Supernova explosions in the interior of the superbubbles appear to be
the most likely cause of this intermittent excess X-ray luminosity.
However, a supernova in the center of a superbubble will not produce
significant excess X-ray emission, as its blast wave expands into hot,
low-density, interior gas and only weakly heats and compresses it.
Two mechanisms have been proposed to enhance the emission from
interior supernovae.  Chu \& Mac Low (1990) show that off-center
supernovae can drive shock waves into the ionized inner edge of the
swept-up shell that can produce the observed X-ray luminosity for a
few thousand years.  Arthur \& Henney (1996) suggested that if the
interior is full of small clumps of unspecified origin, they would
cause enhanced emission when shocked by a supernova blast wave.  

A further twist to the tale comes from the discovery by Oey (1996)
that X-ray bright superbubbles show faster expansion velocities than
can be explained by the mechanical luminosity expected from the
observed interior stellar population (including any supernovae
expected to have already occurred for older clusters), while X-ray dim
superbubbles appear to have dynamics consistent with their stellar
populations. 

One explanation for the discrepancy between the observed dynamics and
stellar population that has been developed is that these rather small
superbubbles (with diameters of order only 100 pc) are not expanding
in a uniform medium, but rather are blowing out of their dense
parental clouds.  \nocite{c97}Comer\'on (1997) performed numerical
models of this process, and \nocite{sf99}Silich \& Franco (1999) showed using
thin-shell models that such a blowout can explain the observed high
velocities.  However, they predict that the fast-expanding bubbles
should be a factor of 3 {\em dimmer} than spherical bubbles rather
than the observed factor of 3--10 {\em brighter}, leaving the puzzle
unsolved. 

All of these models make very simple assumptions about the medium that
the superbubbles are expanding into, however: either that it is
homogeneous, or that there is a slab of higher density in a homogenous
lower-density medium.  However H~{\sc i} observations of the LMC show
that the gas is highly structured, with the cold gas being confined
between a foam of bubbles \nocite{k98}(Kim, et al.\ 1998).
This structure has been reproduced in three-dimensional computational
models that include disk stratification and supernova driving
\nocite{a00,k99a,k99b}(Korpi et al.\ 1999a,b; Avillez 2000).  An interesting
question for the future is whether superbubble expansion in such a
medium can reproduce the observational constraints that have prevented
a complete theoretical model to date.

\section{Summary}

I have here reviewed our current understanding of the interiors of
stellar wind bubbles and superbubbles.  Although there is a detailed
theoretical description and plenty of observational constraints, I
find that there are some important open issues that have not yet been
pinned down.

In the case of stellar wind bubbles, the existence and strength of
evaporative conduction will play a crucial role in determining their
interior thermal history, perhaps helping to explain the evidence for
cold main-sequence bubble interiors.  Corrugation and clumping of the
swept-up shell due to Vishniac instabilities and a turbulent ambient
medium may also contribute to the cooling of the interior.  The
effects of non-equilibrium ionization, as well as enhanced interior
metal abundances, must be taken into account in a dynamical model that
includes the effects of post-main sequence evolution in order to
understand whether observations of Wolf-Rayet bubble X-ray emission
are really consistent with theory or not.  There appear to be several
hints pointing to enhanced cooling due to these various effects, but 
solid explanation and observational confirmation is still lacking.

In the case of superbubbles, the set of constraints posed by
observations of central stars, shell dynamics, and X-ray luminosities
also does not appear to be fit by any single theoretical model.  The
central problem might be best summarized by asking, ``Why are fast
bubbles X-ray bright?''  The best explanation of fast bubbles as
blowouts from their natal molecular clouds would predict that they are
X-ray dim rather than X-ray bright, while the best explanation of
X-ray bright bubbles as due to off-center supernovae probably cannot
explain the strong observed acceleration of the shells (although this
point has not been proven with quantitative work so far as I know).
The assumption of uniformity or simple structure to the surrounding
medium may well be to blame, however, as the actual medium with which
superbubbles interact is strongly structured by previous generations
of OB stars.  Simulations incorporating this structure need to be used
to understand whether that can explain the observed dynamics and X-ray
emission. 

\acknowledgements  I thank the organizers for their invitation and
partial support of my attendance at this meeting.

\end{document}